# Light-Emitting Diodes in the Solid-State Lighting Systems

Amelia Carolina Sparavigna[1]

[1]Department of Applied Science and Technology, Politecnico di Torino, Torino, Italy

**Abstract:** Red and green light-emitting diodes (LEDs) had been produced for several decades before blue emitting diodes, suitable for lighting applications, were widely available. Today, we have the possibility of combining the three fundamental colours to have a bright white light. And therefore, a new form of lighting, the solid-state lighting, has now become a reality. Here we discuss LEDs and some of their applications in displays and lamps.

## 1. Introduction

The Royal Swedish Academy of Sciences has decided to award the Nobel Prize in Physics for 2014 to Isamu Akasaki, Hiroshi Amano and Shuji Nakamura, because of their fundamental works on blue light-emitting diodes (LEDs). In his will, Alfred Nobel dictated that the winners were persons who had conferred great benefit to humankind. In the spirit of the founder then, this year the Prize is awarding the creation of a new light source, which is more energy-efficient and environment-friendly of the previous ones, where a bright white light is obtained by using a combination of red, green and blue lights emitted by diodes [1].

The Japanese Nobel Laureates developed blue light from semiconductors in the early 1990s. Red and green emitting diodes had been already produced and used from several years, but without a blue emitting diode, the white light was impossible. Shuji Nakamura, working for a small company called Nichia Chemical Industries, was responsible for developing some manufacturing processes that brought first truly bright blue LEDs to market. However [2,3], this achievement could not have been done without the work of Akasaki and Amano, who deeply studied gallium nitride (GaN), the material used to obtain the blue LEDs, after the Radio Corporation of America, RCA, abandoned it in the early 1970s. At a laboratory of Nagoya University, Akasaki and his doctoral student Amano developed a way to make high-quality gallium nitride, producing a buffer layer of aluminum nitride on top of a sapphire substrate, before growing it. The buffer layer better matched the crystal structures, and in this manner, the flaws occurring in the production of GaN-based LEDs were strongly reduced. This work, published in 1986 [4], attracted the attention of Nakamura. In 1989, the Akasaki's team announced the first p-n junction LED, nearly 100 times brighter of the previous LEDs [2]. Nakamura's group followed with a lower-cost doping method and several innovations that increased quality and brightness, so that Nakamura's company was the first to begin selling blue p-n junction LEDs [2].

The gallium nitride is then the semiconducting material, which is at the center of a new industry, that of the solid-state lighting (SSL), where solid-state devices are producing light. For many decades, chemists, physicists, and engineers in the United States and Japan struggled with the persistent material challenges GaN was presenting, to meet the requests of potential markets in lighting and consumer electronics [3]. Competition among firms led to the shaping of gallium nitride into a critical material for the manufacture of light emitting diodes [3]. Today, the consequence of all these researches is the advent of LED lamps, which are long lasting and efficient alternatives to older light sources.

Many consider the LED lamps so revolutionary, as it was the advent of the incandescent lamps at the beginning of the XX century. For this reason, this new technology deserves some discussions; let us start from the light-emitting diodes, their history and physics, and continue with some applications of SSL illumination system.

## 2. History of LEDs

The incandescent light bulbs had lit up our life since the beginning of the Twentieth Century, when they superseded the first electric lamps, which were the carbon-arc lamps [5]. In the incandescent bulbs, a large part of the power consumed is converted into heat rather than visible light [6]. Since other electrical light sources are more effective, the incandescent bulbs raise some financial and ecological concerns [7]. The first alternative to the incandescent light bulb was the high-efficiency compact fluorescent lamp, or CFL [8]. However, CFLs have problems about the inclusion of mercury in the design and have, sometimes, a color quite different from that of incandescent lamps. Today, the solution for a better lighting system seems to be the use of LEDs.

The light-emitting diodes, LEDs, have been known for many years. They are based on the electroluminescence, which is the production of light by the flow of electrons. This phenomenon was discovered in 1907 by H.J. Round of Marconi Laboratories, who used a crystal of silicon carbide (SiC) and a cat's-whisker detector to show it [9,10]. Oleg Losev, a Russian scientist, is the first who created a LED and reported a detailed study of it in a publication of 1927 [10]. In fact, in 1920s, Losev studied the light emission from zinc oxide and silicon carbide crystal rectifier diodes, which were used in radio receivers. He recognized the cold, non-thermal, nature of this emission, concluding that LED emission was related to the diode action [10]. In 1929, in introducing a patent, Losev wrote that LEDs could be used as detectors in optical relays for fast telegraphic and telephone communications, transmission of images and other applications, understanding the potential of LEDs for telecommunications [10].

In 1955, Rubin Braunstein, of the Radio Corporation of America (RCA), reported on infrared emission from gallium arsenide (GaAs) and other semiconductor alloys [11]. After, in 1957, Braunstein further demonstrated that such devices could be used for non-radio communication across a short distance. In 1962, the Texas Instruments began a project to manufacture infrared diodes, and in the October of that year, announced the first LED commercial product, which employed a pure GaAs crystal to emit a 900 nm output [12]. During the same year, Nick Holonyak Jr., while working at General Electric Company [13], developed the first commercial visible red LED. A former graduate student of Holonyak, George Craford, invented the first yellow LED and improved the brightness of red and red-orange LEDs by a factor of ten in 1972 [14].

The first applications of LEDs were in the replacements for incandescent and neon indicator lamps and in the seven-segment displays (SSD), first in expensive equipment such as laboratory and electronics test equipment, then later in TVs, radios, telephones, calculators and watches [15]. It seems, as told in [15], that seven-segment displays can be found in patents as early as 1908. These displays did not achieve widespread use until the advent of LEDs; today, however, for many applications, dot-matrix liquid crystal displays (LCDs) have largely superseded the LED displays [16].

In fact, visible and infrared LEDs were extremely costly until, in 1968, Monsanto Company organized a mass-production of visible LEDs for indicators, using gallium arsenide phosphide (GaAsP) [17]. Hewlett Packard (HP) introduced these LEDs, which were bright enough only for use as indicators, as the light output was not enough to illuminate an area [12]. Later, other colors became available (Figure 1), and LED materials more advanced, so that the light output rose, while maintaining efficiency at acceptable levels [12]. At last, in 1994, Nakamura produced the first high-brightness blue LED based on indium gallium nitride (InGaN) [18]. Today we have LEDs of several colours in digital clocks, flashlights and traffic signals. Only recently, however, companies started manufacturing them in standard, replacement-size light bulb form, so that the LED lamps can be used into a lamp holder like the old incandescent bulbs.

### 3. LED Physics

A LED is a component of the solid-state lighting (SSL) technology. Instead of emitting light from an incandescent wire or from a gas, a SSL device emits light from the bulk of a semiconducting material by electroluminescence. This is the result of radiative recombination of electrons and holes, where the excited electrons release energy as photons. In LEDs, prior to recombination, electrons and holes are separated by doping the material to form a p-n junction, made of positively and negatively charged components. The positive layer has holes, that is, vacancies of electrons; the negative layer has free electrons than can move in it. When the flow of electrons from the negative to the positive layer is activated, the electrons emit light as they have a radiative recombination with holes. The wavelength of the light emitted, and thus its color, depends on the band gap energy, that is, the energy difference between the top of the valence band and the bottom of the conduction band, possessed by the materials giving the p-n junction. The emission of photons is also depending on the type, direct or indirect, of the band gap. Among the materials used for LEDs, we find those with a direct band gap with energies corresponding to light ranging from near-infrared to near-ultraviolet light [19], and those, such as gallium phosphide (GaP), having indirect band gaps. Let us consider that GaP is used in the manufacture of low-cost red, orange, and green LEDs, with low to medium brightness, since the 1960s.

Semiconductors can have direct or indirect band gaps (see Figure 2). A certain crystal momentum, and its corresponding wave k-vector in the Brillouin Zone, characterizes the minimum energy state of the conduction band; another k-vector characterizes the maximum energy state of the valence band. If these two k-vectors are the same, we have a direct gap. If they are different, there is an indirect gap. When the band gap is direct, an electron can recombine with a hole of the same momentum, directly emitting a photon. This photon has a relative large wavelength, and therefore a negligible wavenumber, when compared to lattice k-vectors. Therefore, the conservation of the total crystal momentum is easily fulfilled. When the gap is indirect, the probability of a photon emission is lower, because electrons must transfer momentum to the phonons of the crystal lattice [20]. Gallium phosphide for instance, with its normal cubic crystal structure having an indirect band gap, is severely limited in its emission efficiency. To increase the emission, GaP nanowires with pure hexagonal crystal structure having a direct band gap had been prepared, with a strong photoluminescence at a wavelength of 594 nm. In these nanowires, the incorporation of aluminum or arsenic can tune the emitted wavelength across an important range of the visible light spectrum. This approach to crystal structure engineering enables new pathways for tailoring the materials properties [21,22].

The active regions of LEDs can be created by epitaxy, in a process that consists essentially by growing a layer of crystal doped with one type of dopant on top of a layer of crystal doped with another type of dopant. Actually, epitaxy is a technology used in the fabrication of integrated crystalline layers of different materials, especially if based on GaN and InGaN, grown on sapphire substrates [23,24]. For what concerns the internal electroluminescence of LEDs, it is high, but, due to the very high refractive indices of the materials used for them (2.5 for GaN and 1.7 for sapphire), the external luminescence is limited [25]. In fact, a large part of the light is reflected at the material/air interface back into the bulk of the chip. Therefore, besides the development of luminescent materials, another important aspect of LEDs production is the light extraction from them. Commonly, the reflection is reduced by using a dome-shaped encapsulation of the diode, so that the emitted rays of light strike the LED surface package perpendicularly. Substrates that are transparent to the emitted wavelength, and backed by a reflective layer, increase the LED efficiency.

**4. LED Packages**
Let us discuss the packaging of LEDs with more details. The fabricated LED chips are mounted in a package that consists of two electrical leads, a transparent optical window for the escape of light and thermal paths for heat dissipation. A typical package for low-power devices is given in the Figure 3 (on the left). The active device is bonded or soldered to the bottom of a cup-like depression, the reflector cup, with one of the lead wires, usually the cathode lead. A bond wire connects the LED top contact to the other lead wire, usually the anode lead. This package is frequently referred to as the 5mm or T1-3/4 package [26]: its encapsulating material has a hemispherical shape to maintain the incidence of the rays of light normal to the encapsulate-air interface. The standard 5mm LED package is not suited to have a sufficient heat transfer to maintain the LED cool during operation. Then, a new package for high-power LEDs has been developed: it is the flip-chip package that we can see in the Figure 3, on the right [26,27].

Today, LEDs have an emission high enough allowing the production of LED lamps, which are nothing else that LEDs assembled into light bulbs to be used in the common lighting fixtures. These lamps have a lifespan and electrical efficiency, which are several times better than incandescent lamps. Even when compared with the fluorescent lamps, LEDs are significantly better. Let us remark that, like the traditional incandescent lamps, LEDs have a full brightness without needing for a warm-up time, which is required by most fluorescent lamps. Since LEDs have not an isotropic emission, the lamps need a specific design, and, because the emission of a single LEDs is less than that of incandescent and compact fluorescent lamps, several LEDs are used to form the lamp. Moreover, LEDs need direct current and then a circuit to convert the alternating current.

**5. Advantages of LEDs**
It is highly probable that in the future many household lighting fixtures will be equipped with LED lamps. A good reason to want them is in their reduced energy use, since LEDs produce light losing far less energy to heat than incandescent lamps. LED lamps are around 80 percent more efficient, and are more efficient than CFLs too. A LED bulb last up to 25,000 hours, compared with 2,000 hours for a standard bulb and 8,000 for a compact fluorescent, which means a life of 17 years if the bulb is on four hours a day [28]. The annual energy use is then reduced and the corresponding annual $CO_2$ emission in atmosphere. Moreover, LED lamps produce the same white light as that of incandescent bulbs.

Today, LED lamps are expensive. These lamps are of course saving money in the long run, because we need to replace them every decade or two, and the electric bill is reduced. However, their upfront cost is prohibitive for people who simply cannot spend several hundred dollars for equipping the fixtures of their houses with such bulbs. Consequently, price is the only real problem with LED light bulbs; but probably, it could change in the next future.

**6. LED Displays**
Today, LEDs are used to create displays, in particular the flat panel displays made of large arrays of light-emitting diodes. Typically, we find them outdoors in store signs, billboards and for the signs on public transport vehicles. The LED panels are of two types: the conventional one using discrete LEDs, and the new one using surface-mounted devices (SMDs). Most outdoor screens have discrete LEDs, which are mounted to form full-color pixels in clusters of red, green, and blue individual diodes, each having its own package. However, most indoors screens are based on the SMD technology, a trend that is now extending to the outdoor market, where SMD pixels are consisting of red, green, and blue diodes, each smaller than a pinhead, set very close together and mounted in a single package (see for instance, Fig.3 top-right).

Discrete LEDs are also known as PTH LEDs, that is, Plating Through Holes LEDs (see Fig.3, top-left). They are welded on the outside of the electronic board, and come into contact with the card using the two feet (anode and cathode) [29]. With the new SMD technology, LEDs are always mounted on the surface of the cards. The conventional PTH LEDs are important for the market of high-brightness LEDs for outdoor video wall applications, because they have good primary optics, and it is difficult to replace them with SMD LEDs and have the same final results [29]. However, the trend is to use SMD technology for outdoor displays too, because it is the same used for the package platform of lighting lamps [29].

For television screens, there are high-brightness diodes able to generate a wide spectrum of colors: these screens are based on organic light-emitting diodes (OLEDs). An organic light-emitting diode is a LED in which the emissive electroluminescent layer is a film of organic compound. This layer is situated between two electrodes, at least one of them being transparent. There are two main families of OLEDs: those based on small molecules and those employing polymers; in this case, they are known as PLEDs, that is, polymeric LEDs. Today, researchers are working to develop white OLED devices for use in solid-state lighting applications [30].

Let us note that we have also LED-backlit LCD displays. They are flat panel displays which use LED backlighting arrays instead of cold cathode fluorescent lamps (CCFLs), which are commonly used in LCDs. LED-backlit LCD TVs use the same TFT LCD (thin film transistor liquid crystal display) technologies as CCFL-backlit LCD TVs. Picture quality is primarily based on TFT LCD technology, independent of backlight type. While not LED displays, these televisions are called "LED TV" by some manufacturers and suppliers [31].

### 7. LED Lamps, Art and Fashion
As lighting systems, the LED lamps are used for art exhibitions and museums. Let us see one remarkable example: they are used for one of the masterpieces of Italian Renaissance, the Sistine Chapel of Vatican Apostolic Palaces in Rome. For the past 500 years, people admired it with the light of sun entering the windows of the Chapel. However, from 1980s, Vatican officials, being concerned about sun radiation could damage the paints, screened the windows and used a system of halogen low-energy lamps to preserve pigments on the walls. Being not the best for admiring the Chapel, on October 29, 2014, halogen lamps had been replaced by a system composed of about 7,000 LEDs, mounted in some fixtures placed around the perimeter of the Chapel. The system was provided by Osram, the German lighting manufacturer [32]. The new lighting is aimed precisely in the same direction the natural light entered the windows, with its fixtures hidden below them. In 2013, thank of Toshiba, Paris's Louvre Museum had LEDs to light, among other works, Leonardo da Vinci's Mona Lisa. Also, the London's National Portrait Gallery, the Amsterdam's Rijksmuseum, and the Boston's Museum of Fine Arts had installed LED lighting systems [33].

Of course, we have also the LED Art, which is a form of light art obtained by means of light-emitting diodes. Some artists use this art for producing temporary pieces in public locations. For instance, colored "LED throwies" can be used to "paint" metallic objects, like public sculptures or road infrastructures, without damaging them [34]. Moreover, LEDs can become glowing accessories for clothing and shoes, endowed with special features such as lighting mode, additional effects and control, in a unique mixture of fashion and technology. In this manner, our shoes for instance, can twinkle when we move.

### 8. Some Current Researches on LEDs
As we have seen, LEDs have several advantages over traditional lighting systems, but, today, their price is relatively high. LEDs are expensive because the manufacturing process used to fabricate the wafers from which LED chips are obtained is difficult. The difficulties are coming from the employed materials, usually gallium nitride for diode layers and sapphire for substrates. To reduce the costs, some manufacturers have proposed using silicon as substrate, and silicon is the material routinely used to fabricate the integrated circuits. This proposal is not recent: silicon had been tried, but large mismatches between its crystal structure and that of GaN create faults that reduce the product's efficacy. New developments promise to overcome these drawbacks [35].

As told previously, LED chips are obtained by deposition of several layers of different materials starting from a given substrate [36]. Multiple thin layers, grown by epitaxy and sandwiched between p- and n-contacts, are usually composing the active region of chips. Normally, these active layers are InGaN/GaN multiple quantum wells (MQWs) [37,38]. A quantum well is a potential well used to confine particles, which are originally free, allowing them to occupy several energy subbands. In such a manner, the indium-rich regions of quantum wells are successfully capturing electron-hole pairs that recombine with emission of light. Because of the presence of a large lattice and thermal mismatch between InGaN and GaN, a strain between the active layers and the underlying n-GaN template exists [38]. For this reason, several solutions are studied to release the residual strain and enhance the electron capture rate of the active region too (see [38] and references therein). Other researches are concerning lattice defects, such as dislocations, and their role as non-radiative recombination sites in nitride materials [39,40].

The above-mentioned studies, such as several other researches, are currently performed to increase the efficiency of InGaN LEDs. However, besides being fundamental for the blue LEDs, InGaN is a very interesting material for solar photovoltaic devices too [41-43]. For this reason, InGaN and InGaN MQWs are also attracting several researches on this subject, and, probably, we will see further increased efficiencies of LEDs accompanied by the same achievements for solar cells.

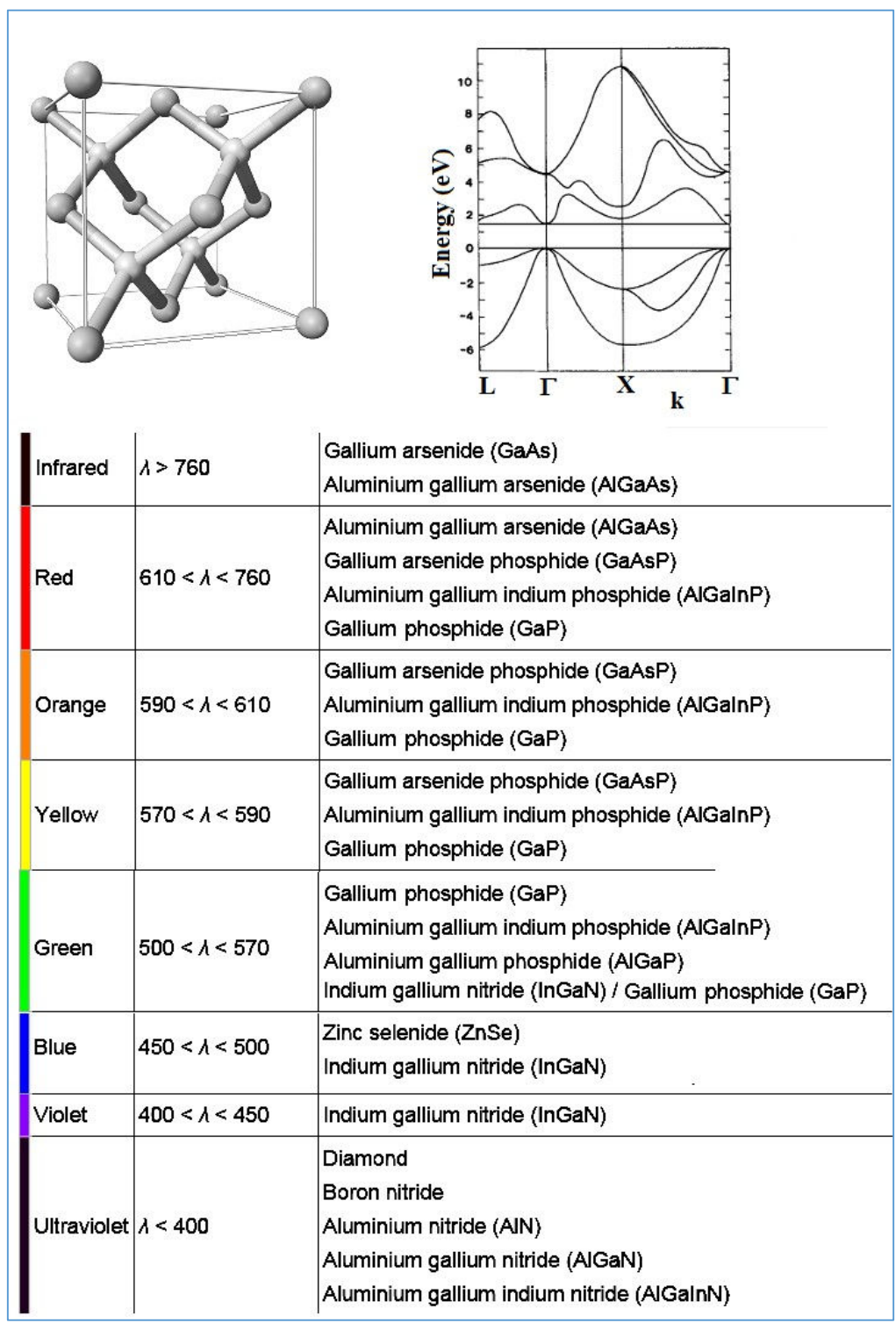

**Figure 1:** LED development began with infrared and red devices made of gallium arsenide (GaAs). Advances in materials science enabled to create devices with shorter wavelengths, emitting light in a variety of colors, using different semiconductors. In the upper part of the image, we see the lattice of GaAs (courtesy Wikipedia) and the energy band structure as a function of the wavenumber in the Brillouin Zone, as calculated by F. Herman and W.E. Spicer, in their paper entitled "Spectral analysis of photoemission yields in GaAs and related crystals", published in the Physical Review of 1968.

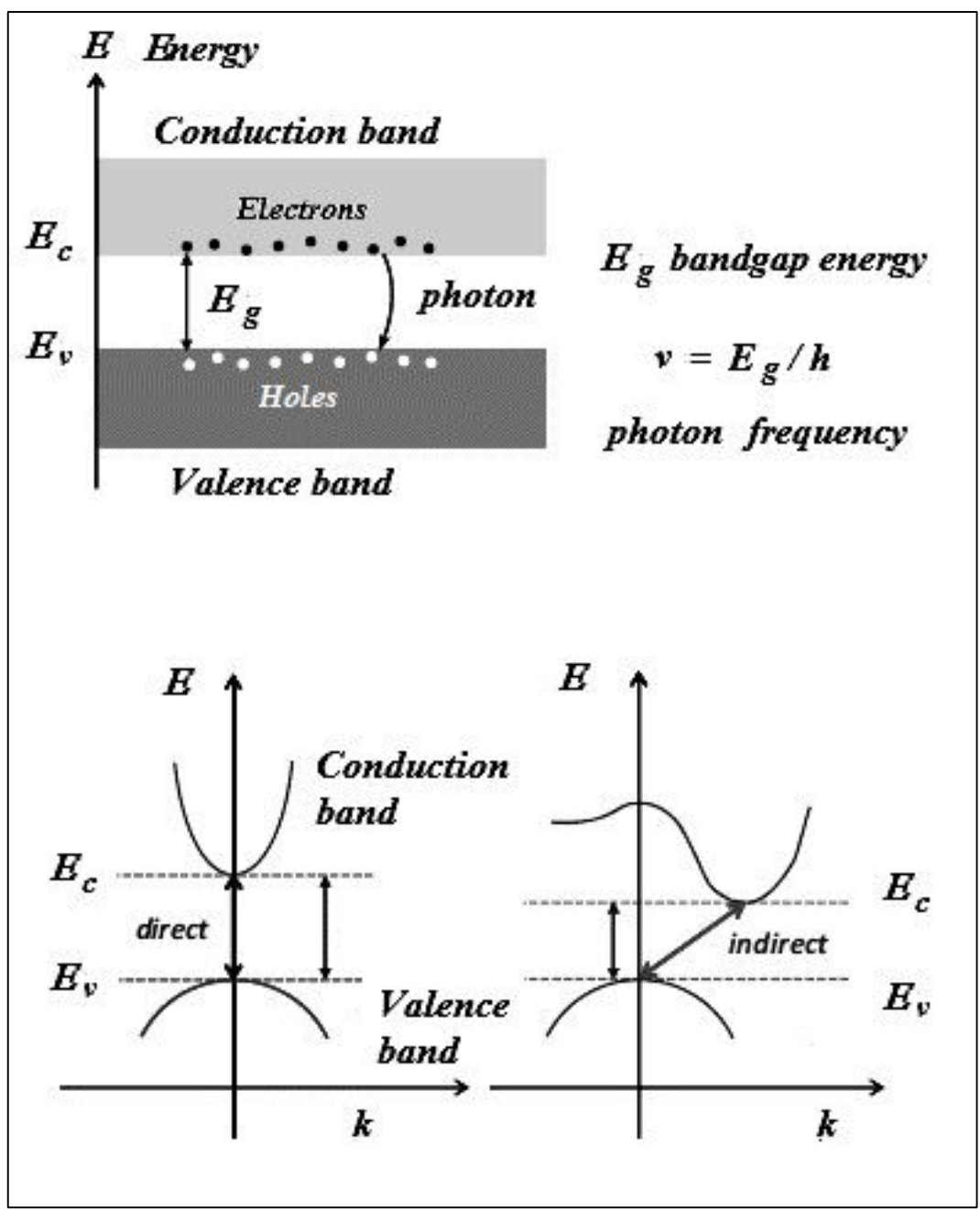

**Figure 2**: The frequency of the emitted light depends on the band-gap energy of p-n junction. This gap exists between the bottom of the conduction band and the top of the valence band. In semiconductors, we can have direct or indirect band gaps. A certain crystal momentum, and its corresponding k-vector in the Brillouin Zone, characterizes the minimum energy state of the conduction band; another k-vector gives the maximum energy state of the valence band. If these two k-vectors are the same, we have a direct gap: if different, there is an indirect gap. Therefore, the band-gap is direct if an electron can recombine with a hole of the same momentum, emitting a photon. This photon has a relatively large wavelength, and therefore a negligible wavenumber, when compared to the k-vectors of electrons. Therefore, the conservation of total crystal momentum is easily fulfilled. When the gap is indirect, the probability of a photon emission is very low and electron must transfer momentum to the phonons of the crystal lattice [20].

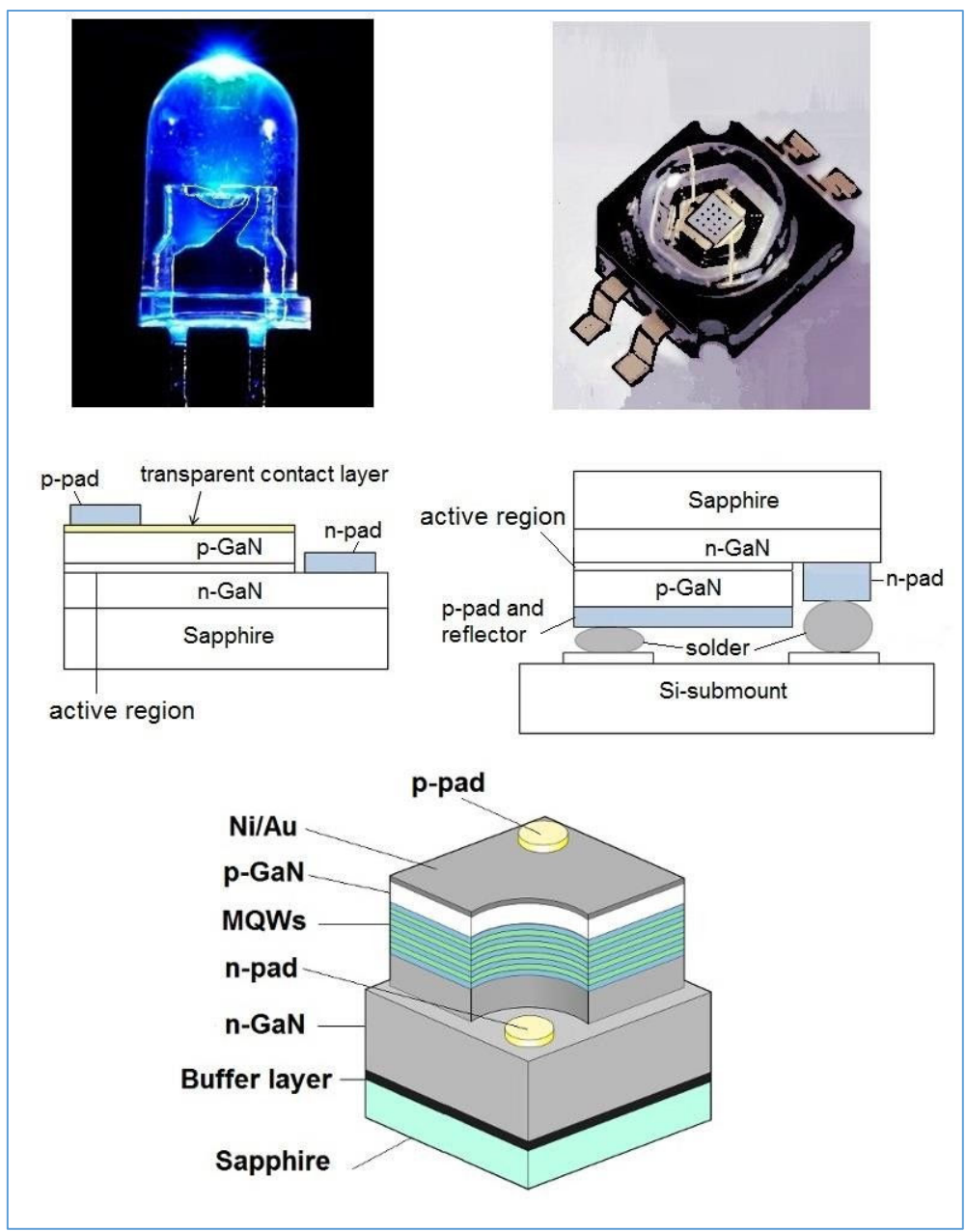

**Figure 3:** The fabricated LED chips are mounted in a package that consists of two electrical leads, a transparent optical window for the escape of light and thermal paths for heat dissipation. Typical package for low-power device is given in the figure on the left. In the upper part we see an image of a LED and in the lower part a schematic representation of the diode, in the case of a GaN device. The active device is bonded or soldered to the bottom of a cup-like depression, the reflector cup, with one of the lead wires, usually the cathode lead. A bond wire connects the LED top contact to the other lead wire, usually the anode lead. This package is frequently referred to as the 5 mm or T1-3/4 package [26]: it is encapsulated in a transparent material having a hemispherical shape to have a normal angle of incidence of the rays of light to the encapsulate-air interface. The standard 5mm LED package is not suited to have a sufficient heat transfer to maintain the LED cool during operation. Therefore, a new package for high-power LED had been developed: it is the flip-chip package that we can see in the image on the right, with its schematic view [26,27]. Let us note that the active region is often made by multi quantum wells (the 3D model is a Wikipedia/Tosaka|Tosa courtesy).